\documentclass[12pt]{amsart}
\usepackage{latexsym,epsfig,amssymb,amsmath,amsthm,color,url,tikz}
\usetikzlibrary{arrows,positioning} 
\usepackage{graphicx}
\usepackage{inputenc}
\usepackage[square,comma,sort&compress]{natbib}
\setcitestyle{numbers,multirow,import}
\usepackage{graphicx}
\allowdisplaybreaks
\numberwithin{equation}{section}
\numberwithin{figure}{section}

\newtheorem{Theorem}{Theorem}[section]
\newtheorem{Lemma}[Theorem]{Lemma}
\newtheorem{Remark}[Theorem]{Remark}
\newtheorem{Example}[Theorem]{Example}
\newtheorem{Proposition}[Theorem]{Proposition}
\newtheorem{Definition}[Theorem]{Definition}
\newtheorem{Corollary}[Theorem]{Corollary}

\newcommand{\Fin}{F_\text{in}}
\newcommand{\ntail}{n_{\text{tail}}}

\newcommand{\deltain}{\delta_\text{in}}
\newcommand{\deltaout}{\delta_\text{out}}

\newcommand{\bthe}{\begin{Theorem}}
\newcommand{\ethe}{\end{Theorem}}

\newcommand{\ble}{\begin{Lemma}}
\newcommand{\ele}{\end{Lemma}}

\newcommand{\bde}{\begin{Definition}}
\newcommand{\ede}{\end{Definition}}

\newcommand{\bco}{\begin{Corollary}}
\newcommand{\eco}{\end{Corollary}}

\newcommand{\bpr}{\begin{Proposition}}
\newcommand{\epr}{\end{Proposition}}

\newcommand{\brem}{\begin{Remark}}
\newcommand{\erem}{\end{Remark}}

\newcommand{\bexam}{\begin{Example}}
\newcommand{\eexam}{\end{Example}}

\newcommand{\beqq}{\begin{equation}}
\newcommand{\eeqq}{\end{equation}}

\newcommand{\beao}{\begin{eqnarray*}}
\newcommand{\eeao}{\end{eqnarray*}\noindent}

\newcommand{\beam}{\begin{eqnarray}}
\newcommand{\eeam}{\end{eqnarray}\noindent}

\newcommand{\barr}{\begin{array}}
\newcommand{\earr}{\end{array}}

\newcommand{\bproof}{\begin{proof}}
\newcommand{\eproof}{\end{proof}}

\newcommand{\new}[1]{{\color{black} #1}}
\newcommand{\sid}[1]{{\color{black} #1}}
\newcommand{\wtd}[1]{{\color{black} #1}}
\newcommand{\phy}[1]{{\color{black} #1}}

\newcommand{\Din}{I_n}

\newcommand{\Nin}{N^{\text{in}}}
\newcommand{\Nout}{N^{\text{out}}}
\newcommand{\pin}{p^{\text{in}}}
\newcommand{\pout}{p^{\text{out}}}
\newcommand{\qin}{q^{\text{in}}}
\newcommand{\qout}{q^{\text{out}}}
\newcommand{\Cin}{C_{\text{in}}}
\newcommand{\Cout}{C_{\text{out}}}
\newcommand{\ain}{\iota_\text{in}}
\newcommand{\aout}{\iota_\text{out}}
\newcommand{\din}{\delta_{\text{in}}}
\newcommand{\dout}{\delta_{\text{out}}}
\newcommand{\hatdin}{\hat{\delta}_{\text{in}}}
\newcommand{\hatdout}{\hat{\delta}_{\text{out}}}

\newcommand\hatain{\hat\iota_{\text{in}}}
\newcommand\hataout{\hat\iota_{\text{out}}}

\newcommand{\pij}{p_{ij}}

\newcommand{\dd}{\mathrm{d}}
\newcommand{\PP}{\textbf{P}}
\newcommand{\EE}{\textbf{E}}
\newcommand{\ind}{\textbf{1}}
\newcommand{\argmax}{\operatornamewithlimits{argmax}}

\newcommand{\convas}{\stackrel{\text{a.s.}}{\longrightarrow}}
\newcommand{\RR}{\mathbb{R}}
\newcommand{\supy}{\sup_{y\ge 1}}

\newcommand{\argmin}{\operatornamewithlimits{argmin}}

\begin{document}
\bibliographystyle{plain}

\title[Extreme Value Estimation]{Are Extreme Value Estimation Methods Useful for
  Network Data?}

\author[P.\ Wan]{Phyllis Wan}
\address{Phyllis Wan\\Department of Statistics\\
1255 Amsterdam Avenue, MC 4690\\
Columbia University\\
New York, NY 10027}
\email{phyllis@stat.columbia.edu}

\author[T.\ Wang]{Tiandong Wang}
\address{Tiandong Wang\\
School of Operations Research and Information Engineering\\
Cornell University \\
Ithaca, NY 14853}
\email{tw398@cornell.edu}

\author[R.A.\ Davis]{Richard A.\ Davis}
\address{Prof.~Richard A. Davis\\Department of Statistics\\
1255 Amsterdam Avenue, MC 4690\\
Room 1004 SSW\\
Columbia University\\
New York, NY 10027}
\email{rdavis@stat.columbia.edu}

\author[S.I.\ Resnick]{Sidney I.\ Resnick}
\address{Prof.~Sidney I.\ Resnick\\
School of Operations Research and Information Engineering\\
Cornell University \\
Ithaca, NY 14853}
\email{sir1@cornell.edu}

\thanks{Research of the four authors was partially supported by Army MURI grant
  W911NF-12-1-0385 to Cornell University.} 



\begin{abstract} {
	Preferential attachment is an appealing edge generating mechanism for modeling social networks.  It provides both an intuitive description of network growth and an explanation for the observed power laws in degree distributions.  However, there are often limitations in fitting parametric network models to data due to the complex nature of real-world networks.  In this paper, we consider a semi-parametric estimation approach by looking at only the nodes with large in- or out-degrees of the network.  This method examines the tail behavior of both the marginal and joint degree distributions and is based on extreme value theory.  We compare it with the existing parametric approaches and demonstrate how it can provide more robust estimates of parameters associated with the network when the data are corrupted or when the model is misspecified.
}
\end{abstract}
\maketitle


\section{Introduction}\label{intro}

\sid{Empirical studies} \cite{kunegis:2013} suggest that the distribution of in- and
out-degrees of the nodes of many social networks have Pareto-like
tails.  The indices of these distributions 
\sid{control the likelihood of nodes with large degrees appearing in the data.}
\sid{Some  social network models, such as preferential attachment,
theoretically exhibit} these heavy-tailed characteristics.  \sid{This
paper estimates heavy tail parameters using 
semi-parametric extreme value (EV) methods and compares such EV
estimates with model-based likelihood methods.}
\sid{The EV} estimates only rely on the upper tail of the
degree distributions so one might expect these estimates \sid{to be}
robust against model \sid{error or data
corruption}.

\sid{Preferential} attachment \sid{(PA)} describes the growth of \sid{a}
  network where edges and nodes are added over time based on
  probabilistic rules \sid{that assume}
 existing nodes with large degrees attract more
  edges. \new{This property is}  attractive \new{for} modeling social
  networks \new{due to} intuitive \sid{appeal} and ability to
  produce power-law networks with degrees \sid{matched to data}
  \cite{durrett:2010b,vanderhofstad:2017, 
   krapivsky:2001,krapivsky:redner:2001,bollobas:borgs:chayes:riordan:2003}. 
Elementary 
 descriptions of the preferential attachment model 
can be found in \cite{easley:kleinberg:2010} while more 
mathematical treatments are available in \cite{durrett:2010b,vanderhofstad:2017,bhamidi:2007}. Also see \cite{kolaczyk:csardi:2014} for a statistical survey of
methods for network data \new{and \cite{MR3707244} for inference for
  an undirected model}. 

The linear preferential attachment model has received most attention.
\sid{M}arginal degree power laws were established in
\cite{krapivsky:2001,krapivsky:redner:2001,bollobas:borgs:chayes:riordan:2003}, while
joint power-law behavior, also know\sid{n} as joint regular variation, was
proved in \cite{resnick:samorodnitsky:towsley:davis:willis:wan:2016,
  resnick:samorodnitsky:2015,wang:resnick:2016} for the directed linear PA
model.  Given observed network data,
\cite{wan:wang:davis:resnick:2017} proposed parametric inference
procedures for the model in two data scenarios. For the case where the
history of network growth is available, the MLE estimators \new{of
model parameters} \phy{were} derived
and shown to be strongly consistent, asymptotically normal and
efficient. For the case where only a snapshot of the network is
available at a single time point, \phy{the} estimators based on
moment methods \wtd{as well as an} approximation to the likelihood \phy{were shown to be}
strongly consistent. The  loss of
efficiency relative to full MLE was surprisingly mild. 

\sid{The drawback of these two methods} is \phy{that} they are model-based and
sensitive to model error. To overcome this lack of robustness, this
paper describes an \sid{EV inference method applied to}
a single
snapshot of a network and \sid{where possible, compares the EV method
  to model-based MLE methods.}
\sid{The EV method is based on estimates of} in- and
out-degree tail indices, $\iota_\wtd{\text{in}}$ and $\iota_\wtd{\text{out}}$,
\sid{using a combination of the Hill estimator \sid{\cite{hill:1975,resnickbook:2007}}
coupled with a minimum
distance thereshold selection method} \cite{clauset:shalizi:newman:2009}.
We also describe
estimation of model parameters using the joint tail distribution of in- and
out-degrees relying on the asymptotic angular measure
\sid{\cite[page  173]{resnickbook:2007}}
density obtained
after standardizing \sid{\cite[page  203]{resnickbook:2007}}
 the data. 

\sid{If the data \phy{are} generated by the linear PA model}, 
the EV estimators can be applied to estimate the parameters of the
model and \sid{compared with MLE estimates} and not surprisingly, the
\sid{EV estimates}  exhibit larger variance.  
\sid{However, if there is model error or data corruption, the EV \phy{estimates}
more than hold their own and we illustrate the comparison in two ways:}
\begin{itemize}
\item The data is corrupted; linear PA data ha\new{ve} edges randomly deleted
  or added. The EV approach 
reliably  recovers the original preferential attachment parameters
while parametric methods degrade considerably.
\item The data comes from a misspecified model, namely a
directed edge superstar model \cite{bhamidi:2015} but is analyzed as
if it comes from the linear PA model.   \sid{The EV method \sid{gives good estimates for
  superstar model
tail} indices and outperforms MLE based on a misspecified linear PA
model if the probability of attaching to the superstar is significant.}
\end{itemize}

The rest of the paper is structured as follows.
Section~\ref{sec:network:ht} formulates the power-law phenomena in
network degree distributions along with joint dependency in the \wtd{in- and out-}
degrees.  We describe two network models which exhibit such
heavy tail properties, the linear \sid{PA and the 
 superstar linear PA models.}  The EV
inference method for networks is \phy{described}
 in Section~\ref{sec:tail} where we discuss its 
 use for estimating the parameters of the linear PA
model. Section~\ref{sec:est}  \sid{gives  EV estimation results for
simulated data from  the linear PA model.}  \sid{Since} the \sid{generating} model is
correctly specified, we use the previous parametric methods as
benchmarks for comparison in Section~\ref{subsec:robust}. 
Section~\ref{sec:perturb} analyzes network data generated from the
linear PA model but \sid{corrupted}  by random edge
addition or deletion.  Pretending ignorance of the perturbation, we
compare the performance of the extreme value method with the MLE and
snapshot methods to recover the original model.  In
Section~\ref{subsec:superstar}, we 
\sid{use our EV inference approach on data from}
the directed superstar
model and attempt to  to recover the tail
properties of the degree distributions.  \sid{A concluding} Section~\ref{sec:discussion}
summarizes the discussion and reasons why EV methods have  their place.
\sid{Appendices give proofs and a fuller discussion of MLE \wtd{and
    the snapshot method} for linear
  PA models abstracted from \cite{wan:wang:davis:resnick:2017}.}
 

\section{Networks and Heavy-Tailed Degree Distributions} \label{sec:network:ht}

\subsection{General discussion.}\label{subsec:gen}
\sid{We begin with a \sid{general} discussion of power laws and networks.}
Let $G(n)=(V(n),E(n))$ denote a directed network, \wtd{where} $V(n)$ is the
set of nodes, $E(n)$ is the set of edges, and $n$ is the number of
edges.  Let $N(n)$ denote the number of nodes in $G(n)$ and $N_n(i,j)$ be the
number of nodes with in-degree $i$ and out-degree $j$. The marginal
counts of nodes with in-degree $i$ and out-degree $j$ are 
given by 
\begin{equation*}\label{e:count}
	\Nin_{i}(n) := \sum_{j=0}^\infty N_n(i,j) \mbox{ \quad and \quad } \Nout_{j}(n) := \sum_{i=0}^\infty N_n(i,j),
\end{equation*}
respectively.  For many network \sid{data sets}, log-log plots of the in- and out-degree distributions, i.e., plots of $\sid{\log i}$ vs.~
$\log\Nin_i(n)$ and $\sid{\log j}$ vs.~$\log\Nout_j(n)$, appear to be
linear \sid{and generative models of network growth seek to reflect this.}
Consider models such that the empirical degree frequency
\new{converges almost surely,}
\begin{equation}\label{pij}
N_n(i,j)/{N(n)} \to p_{ij}, \quad (n\to\infty)
\end{equation}
\wtd{where $p_{ij}$ is} a \sid{bivariate} probability mass
function \wtd{(pmf)}.  The network  exhibits power-law behavior
if 
\begin{align}
\pin_i &:= \sum_{j=0}^\infty \pij \sim \Cin i^{-(1+\ain)}\mbox{ as
         }i\to\infty, \label{eq:pin:pl}\\ 
\pout_j &:= \sum_{i=0}^\infty \pij \sim \Cout j^{-(1+\aout)}\mbox{ as }j\to\infty,\label{eq:pout:pl}
\end{align}
for some positive constants $\Cin,\Cout$.  Let $(I,O)$ be a \sid{fictitious}
 random vector with joint \wtd{pmf} $p_{ij}$, then
\begin{align*}
\PP(I\ge i)&\sim \Cin(1+\ain)^{-1} \cdot i^{-\ain}\mbox{ as }i\to\infty,\\
\PP(O\ge j) &\sim \Cout(1+\aout)^{-1} \cdot j^{-\aout}\mbox{ as }j\to\infty.
\end{align*}

In the linear PA model, the joint distribution of $(I,O)$ satisfies
 non-standard regular variation.   Let $\mathbb{M}(\RR^2_+\setminus \{\boldsymbol 0\})$ be the set of Borel measures on $\RR^2_+\setminus \{\boldsymbol 0\}$ that are finite on sets bounded away from the origin. 
Then $(I,O)$ is {\it non-standard regularly varying} on
$\RR^2_+\setminus \{\boldsymbol 0\}$ \sid{means}
 that \sid{as $t\to\infty$,}
\beqq\label{MRV}
t\PP\left[\left(\frac{I}{t^{1/\iota_\wtd{\text{in}}}},\frac{O}{t^{1/\iota_\wtd{\text{out}}}}\right)\in\cdot\right]\rightarrow \nu(\cdot),\quad \mbox{in }\mathbb{M}(\RR^2_+\setminus \{\boldsymbol 0\}),
\eeqq
where  $\nu(\cdot) \in\mathbb{M}(\RR^2_+\setminus \{\boldsymbol 0\})$
is called the limit or tail measure
\sid{\cite{lindskog:resnick:roy:2014, das:mitra:resnick:2013,
    hult:lindskog:2006a}.}   
Using the power transformation
$I\mapsto I^a$
with {$a = \ain/\aout$}, 
the vector $(I^a,O)$ becomes standard regularly varying, i.e.,
\beqq \label{stdzRV}
t\PP\left[\left(\frac{ I^a}{t^{1/\aout}},\frac{O}{t^{1/\aout}}\right)\in\cdot\right]\rightarrow \tilde{\nu}(\cdot),\quad \mbox{in }\,\mathbb{M}(\RR^2_+\setminus \{\boldsymbol 0\}),
\eeqq 
where $\tilde{\nu}=\nu\circ T^{-1}$ with $T(x,y)=(x^a, y)$.
With this standardization, the transformed measure $\tilde\nu$ is directly estimable from data \citep{resnickbook:2007}.


In the following we describe two classes of preferential attachment models that generate networks with power-law degree distributions.

\subsection{The linear preferential attachment (linear PA) model.}

The directed linear \wtd{PA} model
\cite{bollobas:borgs:chayes:riordan:2003,krapivsky:redner:2001}
constructs a growing {sequence of} directed random graph{s} $G(n)$'s
whose 
dynamics depend on five nonnegative {parameters}
$\alpha, 
\beta, \gamma$, $\din$ and $\dout$, where $\alpha+\beta+\gamma=1$ and $\din,\dout >0$. To avoid degenerate situations,
assume  that each of the numbers $\alpha, 
\beta, \gamma$ is strictly smaller than 1. 

We start with an arbitrary initial finite directed
graph $G({n_0})$ with at least one node and $n_0$ edges. Given an
existing graph $G(n-1)$, a new graph $G(n)$ is obtained by adding a
single edge to $G(n-1)$
\sid{so that}  the graph $G(n)$ contains $n$ edges for all $n\ge n_0$.
Let $I_n(v)$ and $O_n(v)$ 
denote the in- and out-degree of $v\in V(n)$ in $G(n)$, {that is, the number of edges pointing into and out of $v$, respectively}. 
{We allow three
scenarios of edge creation, which are activated by flipping a
3-sided coin with probabilities
$\alpha,\beta$ and $\gamma$.} More formally,  {let $\{J_n, n>n_0\}$ be} an iid sequence
 of {trinomial} random
variables  with cells labelled $1,2,3$ and cell
probabilities $\alpha,\beta,\gamma$. 
Then the graph $G(n)$ is
obtained from  $G(n-1)$ as follows.

\tikzset{
    >=stealth',
    punkt/.style={
           rectangle,
           rounded corners,
           draw=black, very thick,
           text width=6.5em,
           minimum height=2em,
           text centered},
    pil/.style={
           ->,
           thick,
           shorten <=2pt,
           shorten >=2pt,}
}
\newsavebox{\mytikzpic}
\begin{lrbox}{\mytikzpic} 
     \begin{tikzpicture}
    \begin{scope}[xshift=0cm,yshift=1cm]
      \node[draw,circle,fill=white] (s1) at (2,0) {$v$};
      \node[draw,circle,fill=gray!30!white] (s2) at (.5,-1.5) {$w$};
      \draw[->] (s1.south west)--(s2.north east){};
      \draw[dashed] (0,-2.2) circle [x radius=2cm, y radius=15mm];
    \end{scope}
    
     \begin{scope}[xshift=5cm,yshift=1cm]
      \node[draw,circle,fill=gray!30!white] (s1) at (.5,-1.5) {$v$};
      \node[draw,circle,fill=gray!30!white] (s2) at (-.5,-2.5) {$w$};
      \draw[->] (s1.south west)--(s2.north east){};
      \draw[dashed] (0,-2.2) circle [x radius=2cm, y radius=15mm];
    \end{scope}
    
     \begin{scope}[xshift=10cm,yshift=1cm]
      \node[draw,circle,fill=white] (s1) at (2,0) {$v$};
      \node[draw,circle,fill=gray!30!white] (s2) at (.5,-1.5) {$w$};
      \draw[->] (s2.north east)--(s1.south west){};
      \draw[dashed] (0,-2.2) circle [x radius=2cm, y radius=15mm];
    \end{scope}
    
     \node at (0,-3.5) {$\alpha$-scheme};
     \node at (5,-3.5) {$\beta$-scheme};
     \node at (10,-3.5) {$\gamma$-scheme};
  \end{tikzpicture}    

\end{lrbox}
  \begin{figure}[h]
    \centering 
    \usebox{\mytikzpic} 
\end{figure} 

\begin{itemize}
\item 
If $J_n=1$ (with probability
$\alpha$),  append to $G(n-1)$ a new node $v\in V(n)\setminus V(n-1)$ and an edge
$(v,w)$ leading
from $v$ to an existing node $w \in V(n-1)$.
Choose the existing node $w\in V(n-1)$  with probability depending
on its in-degree in $G(n-1)$:
\beqq \label{eq:probIn}
\PP[\text{choose $w\in V(n-1)$}] = \frac{I_{n-1}(w)+\din}{n-1+\din N(n-1)} \,.
\eeqq
\item If $J_n=2$ (with probability $\beta$), add a directed edge
$(v,w) $ to $E({n-1})$ with $v\in V(n-1)=V(n) $ and $w\in V(n-1)=V(n) $ and 
 the existing nodes $v,w$ are chosen independently from the nodes of $G(n-1)$ with
 probabilities 
\beqq \label{eq:probInOut}
\PP[\text{choose $(v,w)$}] = \Bigl(\frac{O_{n-1}(v)+\dout}{n-1+\dout N(n-1)}\Bigr)\Bigl(
 \frac{I_{n-1}(w)+\din}{n-1+\din N(n-1)}\Bigr).
\eeqq
\item  If $J_n=3$ (with probability
$\gamma$),  append to $G(n-1)$ a new node ${v} \in V(n)\setminus V(n-1)$ and an edge $({w,v)}$ leading
from  the existing node ${w}\in V(n-1)$  to the new node ${v}$. Choose
the existing node $w \in V(n-1)$ with probability
\beqq \label{eq:probOut}
\PP[\text{choose }w \in V(n-1)] =
\frac{
O_{n-1}(w)+\delta_{\text{out}}}
{n-1+\delta_{\text{out}}N(n-1)}\,. 
\eeqq
\end{itemize}
{For convenience we call these scenarios the $\alpha$-, $\beta$- and 
$\gamma$-schemes.}
Note that this construction allows for the possibility of {multiple edges between two nodes and self loops.}
This linear preferential attachment \new{model} can be simulated efficiently using
the method {described} in
\cite[Algorithm~1]{wan:wang:davis:resnick:2017}  and linked to
\url{http://www.orie.cornell.edu/orie/research/groups/multheavytail/software.cfm}.

It is shown {in 
\cite{resnick:samorodnitsky:towsley:davis:willis:wan:2016,resnick:samorodnitsky:2015,
wang:resnick:2016} that} the empirical degree distribution 
$$
	\frac{N_n(i,j)}{N(n)} \convas p_{ij},
$$
and the marginals satisfy \eqref{eq:pin:pl} and \eqref{eq:pout:pl}, 
where the tail indices are
\beqq\label{c1c2}
\ain =: \frac{1+\din(\alpha+\gamma)}{\alpha+\beta},\quad\text{and}\quad
\aout =: \frac{1+\dout(\alpha+\gamma)}{\beta+\gamma}.
\eeqq
Furthermore, the joint regular variation condition \eqref{stdzRV} is
satisfied by the limit degree distribution and the limit measure
\cite{resnick:samorodnitsky:towsley:davis:willis:wan:2016} or its
density \cite{wang:resnick:2016}
can be explicitly derived.
We shall use this property for parameter estimation in Section~\ref{sec:tail}.

\subsection{The superstar linear \wtd{PA} model.}
{The key feature of the superstar linear PA model that distinguishes
  it from the standard linear PA model is the existence of a superstar
  node, to which a large proportion of nodes attach.} 
A new parameter $p$ {represents the attachment
  probability}. The $\alpha$-, $\beta$- and $\gamma$-schemes of the
linear PA model are still in action.  {However, for the $\alpha$- and
  $\beta$-schemes, an outgoing edge will attach to the superstar node
  with probability $p$, while \sid{with  probability $1-p$}  it will attach to a non-superstar node 
according to the original linear PA rules.

{For simplicity, the network is initialized with two nodes
$V(1)=\{0,1\}$ where node $0$ is the superstar node. \wtd{We assume at the first step,} there
is an edge pointing from $1\to 0$ so $E_1=\{(1,0)\}$.}
Again each graph $G(n)$ contains $n$ edges for all $n\ge1$.
Let 
\[
V^0(n) := V(n)\setminus \{0\}, \quad\text{and}\quad 
E^0(n) := E(n)\setminus\{(u, 0): u\in V^0(n)\},
\]
{so that $E^0(n)$ is the set of edges in $G(n)$ that do not point to the superstar.} Let $|V^0(n)|$ and $|E^0(n)|$ denote the number of nodes and edges in the non-superstar subgraph of $G(n)$, respectively.

The model is specified through the parameter set $(p,\alpha,\beta,\gamma,\din,\dout)$.  Let $\{B_n: n\ge 1\}$ be another iid sequence of Bernoulli random variables where
$$
\PP(B_n = 1) = p = 1-\PP(B_n=0).
$$
The Markovian graph evolution from $G(n-1)$ to $G(n)$ is modified from the linear PA model as follows. 
\begin{itemize}
\item
If $J_n=1$ (with probability $\alpha$),  append to $G(n-1)$ a new node $v\in V(n)\setminus V(n-1)$ and an edge
$(v,w)$ leading from $v$ to an existing node $w$.
\begin{itemize}
\item If $B_n=1$ (with probability $p$), $w=0$,  the superstar node;
\item If $B_n=0$ (with probability $1-p$), $w\in V^0(n-1)$ is chosen according to the linear PA rule \eqref{eq:probIn} applied to $(V^0(n-1),E^0(n-1))$.
\end{itemize}
\item If $J_n=2$ (with probability $\beta$), add a directed edge
$(v,w)$ to $E({n-1})$ where
\begin{itemize}
\item If $B_n=1$ (with probability $p$), $v=0$ and $w\in V^0(n-1)=V^0(n)$ is chosen with probability \eqref{eq:probIn} applied to $(V^0(n-1),E^0(n-1))$;
\item If $B_n=0$ (with probability $1-p$), $v,w\in V^0(n-1)=V^0(n)$ are chosen with probability \eqref{eq:probInOut} applied to $(V^0(n-1),E^0(n-1))$.
\end{itemize}
\item If $J_n=3$ (with probability
$\gamma$),  append to $G(n-1)$ a new node $w\in V^0(n)\setminus V^0(n-1)$ and an edge $(v,w)$ leading
from the existing node $v\in V^0(n-1)$  to $w$, where $v\in V^0(n-1)$ is chosen with probability \eqref{eq:probOut} applied to $(V^0(n-1),E^0(n-1))$.
\end{itemize}

\wtd{If we use} $\Nin_i(n)$ and $\Nout_j(n)$ \wtd{to denote} the 
number of {\it non\/}-superstar nodes that have in-degree $i$ and
  out-degree $j$, respectively, then
 Theorem~\ref{thm:superstar} shows that
$(\Nin_i(n)/n, \Nout_j(n)/n) \to (\qin_i,\qout_j) $  almost surely 
\sid{where the limits} are deterministic constants \sid{that decay
  like power laws.}
\begin{Theorem}\label{thm:superstar}
Let $(\Nin_i(n), \Nout_j(n))$ be the in- and out-degree counts of {the} non-superstar nodes {of the superstar model}. There exists constants
$\qin_i$ and $\qout_j$ such that as $n\to\infty$,
\[
\frac{\Nin_i(n)}{n} \convas \qin_i,\qquad \frac{\Nout_j(n)}{n} \convas \qout_j.
\]
\wtd{Moreover,}
\begin{enumerate}
\item[(i)] As $i\to\infty$,
\beqq\label{power-in}
\qin_i \sim C'_\text{in}\, i^{-(1+\iota_{\text{in}})}, 
\eeqq
where $C'_\text{in}$ is a positive constant and    
\beqq\label{iotain}
\ain :=
\frac{1-(\alpha+\beta)p+\din(\alpha+\gamma)}{(\alpha+\beta)(1-p)}.
\eeqq
\item[(ii)] As $j\to\infty$,
\beqq\label{power-out}
\qout_j \sim C'_\text{out}\, j^{-(1+\iota_{\text{out}})},
\eeqq
where $C'_\text{out}$ is a positive constant and 
\beqq\label{iotaout}
\aout:= \frac{1+\dout(\alpha+\gamma)}{\beta+\gamma}. 
\eeqq
\end{enumerate}

\end{Theorem}
{The proof of Theorem~\ref{thm:superstar} is provided in Appendix~\ref{subsec:proof:superstar}.}


\section{Estimation {Using Extreme Value Theory}}\label{sec:tail}

In this section, we consider \sid{network parameter estimation}
using extreme value theory.  Given a graph $G(n)$ at a fixed
timestamp, the data available for estimates are the in- and
out-degrees for each node denoted by $(I_n(v),O_n (v))$,
$v=1,\ldots,N(n)$.  Let $F_n(\cdot)$ \wtd{be} the empirical distribution
of this data on $\mathbb{N}\times \mathbb{N}$. Then from
\eqref{pij},  almost surely $F_n$  converges weakly to a limit
distribution $F$ on $\mathbb{N}\times \mathbb{N}$ which is the measure
corresponding to the mass function $\{p_{ij}\}$. Let
$\epsilon_{(i,j)}(\cdot)$ be the Dirac measure concentrating on
$(i,j)$ and we have from  \eqref{pij},
\begin{equation}\label{e:FnF}
F_n(\cdot) =\frac{1}{N(n) }\sum_{v=1}^{N(n)} \epsilon_{(I_n(v),O_n(v) )} \phy{(\cdot)}
=\sum_{i,j} \frac{N_n(i,j)}{N(n)} \epsilon_{(i,j)}(\cdot)\phy{\,\overset{w}\to\,}
\sum_{i,j} p_{ij} \epsilon_{(i,j)}(\cdot)=:F(\cdot).
\end{equation}

\subsection{Estimating tail indices; Hill estimation.}\label{subsubsec:clauset}

We review tail index estimation of $\ain$ ($\aout$ is similar) using
the Hill estimator \cite{hill:1975,resnickbook:2007} applied to
in-degree data $I_n(v)$, $v=1,\ldots,N(n)$. 
From \eqref{eq:pin:pl}, the marginal of $F$, called $\Fin$ is
regularly varying with index $-\ain$.
From Karamata's theorem 
$\ain^{-1}$ can be expressed as a function of $\Fin$ \cite[page
69]{dehaan:ferreira:2006}, 
\beqq \label{eq:ain:limit}
	\ain^{-1} = \lim_{t\to\infty} \frac{\int_t^\infty(\log(u)-\log(t))F_\text{in}(du)}{1-F_\text{in}(t)}.
\eeqq
The Hill estimator of
$\ain^{-1}$  replaces $F_\text{in} (\cdot)$
with the marginal of the empirical distribution in \eqref{e:FnF} of
in-degrees, called $F_{\text{in},n}$, and
$t$ with $I_{(k_n+1)}$ in \eqref{eq:ain:limit}.
Let $I_{(1)} \ge \ldots \ge I_{(N(n))}$ be the decreasing order
statistics of $I_n(v)$, $v=1,\ldots,N(n)$. The resulting
estimator is 
\beao 
	\hatain^{-1}(k_n) &=& \frac{\int_{I_{(k_n+1)}}^\infty (\log(u) - \log(I_{(k_n+1)})) F_{\text{in},n}(du)}{k_n/N(n)} \\
	&=& \frac{1}{k_n} \sum_{j=1}^{k_n} (\log(I_{(j)}) - \log(I_{(k_n+1)})).
\eeao
With iid data, if we assume
$k_n\to\infty$ and $k_n/N(n)\to0$, then the Hill estimator is
consistent. Of course, our network data is not iid but Hill estimation
still works in practice. Consistency for an undirected graph is proven 
in \cite{wang:resnick:2017} but for directed graphs, this is an
unresolved issue.

To select $k_n$ in practice, \cite{clauset:shalizi:newman:2009} proposed computing
 the Kolmogorov-Smirnov (KS) distance between the
empirical distribution of the upper $k$ observations and the power-law
distribution with index $\hatain(k)$:
\[
D_{k}:=\supy \left|\frac{1}{k} \sum_{j=1}^{k}{\bf1}_{\{I_{(j)}/I_{(k+1)}>y\}}-y^{-\hatain(k)}\right|, \quad 1\le k\le n-1.
\]
Then the optimal $k^*$ is  the one that minimizes the KS distance
$$
k^* := \argmin_{1\le k\le n} D_{k},
$$
and the tail index is estimated by $\hatain(k^*)$. This estimator performs well if the thresholded portion 
comes from a Pareto tail and also seems effective in a
  variety of non-iid scenarios.  It is widely used by
data repositories of large network datasets such as KONECT (\url{http://konect.uni-koblenz.de/}) \cite{kunegis:2013} {and is realized in} the R-package {\it poweRlaw\/} \cite{gillespie:2015}.

}


{We refer to the above procedure as the {\it minimum distance method} in estimating $\ain,\aout$ for network data.} There are two issues when applying this method. First, the data is node-based and not
collected from independent repeated
sampling. Secondly,  degree counts are discrete {and} do
not exactly comply with the Pareto assumption made in the minimum
distance method. Our analysis shows that even if we ignore these two
issues, the tail estimates are still reasonably good.

\subsection{Estimating dependency between in- and out-degrees}
If the limiting random vector $(I,O)\sim F$ 
\sid{corresponding to  $p_{ij}$ in \eqref{pij}}
is jointly regularly varying and
satisfies \eqref{stdzRV}, we may apply a polar coordinate
transformation, for example, with the $L_2$-norm,  
$$
(I^a, O)\mapsto (\sqrt{I^{2a}+O^2},\arctan(O/I^a)) := (R,T),
$$
where $a=\ain/\aout$.
Then, with respect to $F$ in \eqref{e:FnF},
 the conditional distribution of \sid{$T$} given $R>r$ converges
weakly (see, for example, \cite[p. 173]{resnickbook:2007}),
$$
	F[T\in\cdot | R>r] \to S(\cdot),\quad r\to\infty,
$$
where $S$ is  the {\it angular measure} and describes the asymptotic
dependence of the standardized pair $(I^a, O)$. 
Since for large $r$,  $F[T\in\cdot | R>r] \approx  S(\cdot)$ and for
large $n$, $F_n
\approx F$, it is plausible that for $r$ and $n$ large
$F_n[T \in \cdot | R>r] \approx S(\cdot)$. Skeptics may check
\cite[p. 307]{resnickbook:2007} for a more precise argument and
recall
 $F_n$ is the empirical measure defined in \eqref{e:FnF}.

Based on  observed degrees $\{(I_n(v),O_n(v));
 v=1,\ldots,N(n)\}$, how does this work in practice? First $a$  
 is replaced by $\hat{a} = \hatain/\hataout$ estimated from
 Section~\ref{subsubsec:clauset}. 
Then the distribution $S$ is estimated via the empirical distribution
of the sample angles $T_n(v):=\arctan(O_n(v)/I_n(v)^{\hat{a}})$ for
which $R_n(v):=\sqrt{I_n(v)^{2\hat{a}}+O_n(v)^2} > r$ exceeds some
large threshold $r$.  This is the
POT (Peaks Over Threshold) methodology {commonly employed} in extreme
value theory \cite{coles:2001}.  

In the cases where the network model is known, $S$ may be specified in
closed form.  For the linear PA model, $S$  has a density that is an
explicit function of  the linear PA parameters
\cite{resnick:samorodnitsky:towsley:davis:willis:wan:2016}.  
\new{After estimating $\ain$ and $\aout$ by the minimum distance
  method, the remaining parameters
can then be estimated 
by an approximate likelihood method that we now explain.}

\subsection{EV estimation for the linear PA
  model}\label{subsec:tailPA} 

From  \eqref{c1c2},
\[
\din = \frac{\ain(\alpha+\beta)-1}{\alpha+\gamma},\quad \dout = \frac{\aout(\beta+\gamma)-1}{\alpha+\gamma},
\]
\phy{so that} \sid{the
linear PA model may be parameterized by}
$\boldsymbol{\theta}=(\alpha,\beta,\gamma, \ain, \aout)$. To construct
the EV estimates,
begin by computing the minimum distance estimates
$\hat\iota^{EV}_{\text{in}},\hat\iota^{EV}_{\text{out}}$ of the in-
and out-degree indices.  The parameter $\beta$, which represents the
proportion of edges connected between existing nodes, \wtd{is} estimated
by $\hat{\beta}^{EV} = 1-N(n)/n$. 

\phy{From} \eqref{stdzRV}, $\arctan(O/I^a)$ given $I^{2a}+O^2>r^2$ converges \new{weakly} as $r\to\infty$ to
{the distribution of} a random variable $\Theta$
\cite[Section~4.1.2]{resnick:samorodnitsky:towsley:davis:willis:wan:2016},
{whose pdf} is given by \sid{($0\leq x\leq \pi/2$)}
\begin{eqnarray}
f_\Theta(x;\alpha,\beta,\gamma,\din,\dout) &\propto &\frac{\gamma}{\din} (\cos x)^{\frac{\din+1}{a}-1} (\sin x)^{\dout-1}
\int_0^\infty t^{\ain+\din+a\dout} e^{-t(\cos x)^{1/a}-t^a\sin x}\dd t\nonumber\\
&&+\frac{\alpha}{\dout} (\cos x)^{\frac{\din}{a}-1} (\sin x)^{\dout}
\int_0^\infty t^{a-1+\ain+\din+a\dout} e^{-t(\cos x)^{1/a}-t^a\sin x}\dd t.\label{densIO}
\end{eqnarray}
By {replacing $\beta,\ain,\aout$ with their estimated values} $\hat\beta^{EV}$, $\hat\iota_{\text{in}}^{EV}$, and $\hat\iota_{\text{out}}^{EV}$ {and setting}
${\gamma} = 1-{\alpha}-\hat\beta^{EV},$
the density \eqref{densIO} can be viewed as a profile likelihood
  function (based on a single observation $x$) of the unknown
  parameter $\alpha$, which we denote by 
$$l(\alpha;x)=
f_\Theta(x;\alpha,\hat\beta^{EV},1-{\alpha}-\hat\beta^{EV},
\hat{\delta}_{\text{in}}^{EV},\hat{\delta}_{\text{out}}^{EV}). 
$$ 
Given the degrees $\bigl((I_n(v),O_n(v)), v\in V(n)\bigr)$,
 $\hat\alpha^{EV}$ {can be} computed by maximizing
the profile likelihood
based on the observations
$(I_n(v),O_n(v))$ for which $R_n(v)> r$
for a large threshold $r$. That is, 
\beqq \label{eq:alpha:solve}
	\hat\alpha^{EV} := \argmax_{\sid{0\leq \alpha \leq 1}}
        \sum_{v=1}^{N(n)}
      \log  l\left(\alpha;\arctan\left(\frac{O_n(v)}{(I_n(v))^{\hat{a}}}\right)\right)
        \mathbf{1}_{\{R_n(v)> r\}},
\eeqq
where $r$ is typically chosen as the 
$(\ntail+1)$-th largest $R_n(v)$'s for a suitable $\ntail$.
This estimation procedure is sometimes referred to as the ``independence estimating equations'' (IEEs) method \cite{chandler:bate:2007,varin:reid:firth:2011}, in which the dependence between observations is ignored. This technique is often used when the joint distribution of the data is unknown or intractable.  Finally, using the constraint, $\alpha+\beta+\gamma=1$, we estimate $\gamma$ by $\hat{\gamma}^{EV}=1-\hat\alpha^{EV}-\hat\beta^{EV}$.

\section{Estimation results} \label{sec:est}

In this section, we demonstrate the estimation of the linear PA and related models through the \wtd{EV} method described in Section~\ref{subsec:tailPA}. In Section~\ref{subsec:robust}, data are \wtd{simulated} from the standard linear PA model and used to estimate the true parameters of the underlying model.  Section~\ref{sec:perturb} considers data generated from the linear PA model but \wtd{corrupted} by random addition or deletion of edges.  Our goal is to estimate the parameters of the original linear PA model.  In Section~\ref{subsec:superstar}, we simulate data from the superstar linear PA model and attempt to use the standard linear PA estimation to recover the degree distributions. 

Throughout the section, the EV method is compared with two parametric
estimation approaches for the linear PA model, \wtd{namely} the MLE and
snapshot (SN) methods, proposed in \cite{wan:wang:davis:resnick:2017}.
For a given network, {when the network history is available, that is,
each edge is marked with the timestamp of its creation}, MLE estimates
are directly computable. {In the case where only a snapshot
  of the network is given at a single point in time (i.e., the
  timestamp information for the creation of the edges \new{is}
  unavailable)}, \new{we have} an estimation procedure combining elements of method
  of moments with an approximation to the likelihood. A
brief summary of the MLE and SN methods is  in
Appendix~\ref{subsec:param_est} and  desirable properties of these
  estimators are in \cite{wan:wang:davis:resnick:2017}.

Note that a main difference between the MLE, SN and EV methods lies
in the amount of data utilized.  The MLE approach requires the entire
growth history of the network while the SN method uses only a single
snapshot of the network.  The EV method, on the other hand, requires
only a subset of a snapshot of the network; only those degree counts
of nodes with large in- or out-degrees.  When the underlying model is
true, MLE is certainly the most efficient, but also hinges on having a
complete data set. As we shall see, in the case where the model is
misspecified, the EV method  provides an attractive and reliable
alternative.

\subsection{Estimation for the linear PA model}\label{subsec:robust}

\subsubsection{{Comparison of EV with MLE and SN}}
\begin{figure}[t]
\includegraphics[scale=0.4]{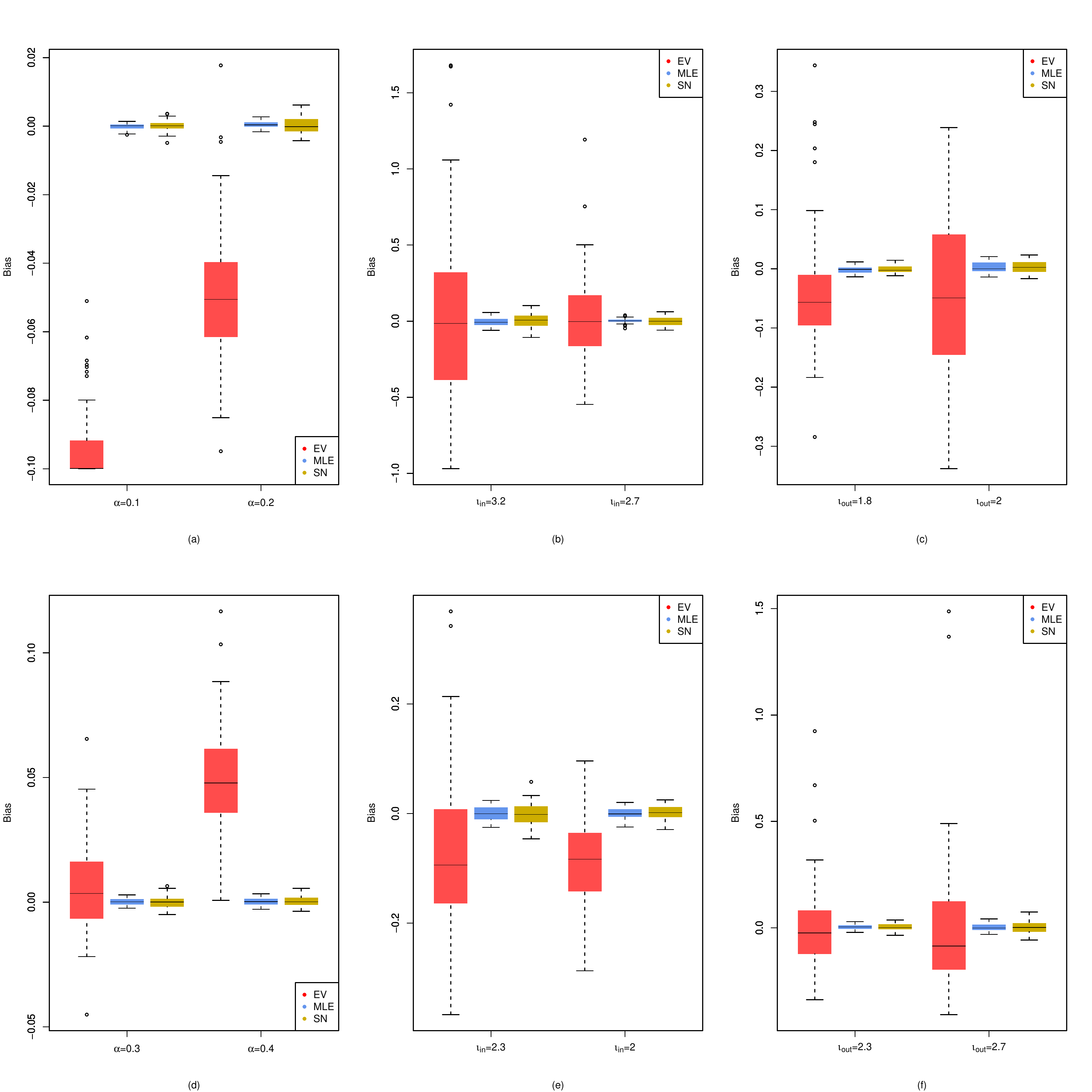}
\caption{Boxplots of biases for estimates of $(\alpha,\ain,\aout)$ using \phy{EV}, MLE and \phy{SN} methods.
Panels (a)--(c) correspond to the case where $\alpha = 0.1, 0.2$ and (d)--(f) are for $\alpha = 0.3, 0.4$, holding
$(\beta, \din, \dout) = (0.4, 1, 1)$ constant.}\label{fig:asy}
\end{figure}
Figure~\ref{fig:asy} presents {biases for estimates of $(\alpha,\ain,\aout)$} using
EV, MLE, and SN methods on data simulated from the linear PA model.

We held $(\beta, \din, \dout) = (0.4, 1, 1)$ constant and varied $\alpha = 0.1, 0.2, 0.3, 0.4$ so that the true values of $\gamma,\ain,\aout$ were also varying. 
For each set of parameter values $(\alpha,\ain, \aout)$, 200 independent replications of a linear PA network with $n=10^5$ edges were simulated and the true values of $(\ain, \aout)$ were computed by \eqref{c1c2}.
We estimated $(\ain, \aout)$ by the minimum distance method 
$(\hat{\iota}^{EV}_\text{in}, \hat{\iota}^{EV}_\text{out})$, MLE and the one-snapshot methods applied to the parametric model (cf.\ Section~\ref{subsec:param_est}), denoted by $(\hat{\iota}^{MLE}_\text{in},\hat{\iota}^{MLE}_\text{out})$ and $(\hat{\iota}^{SN}_\text{in},\hat{\iota}^{SN}_\text{out})$, respectively. With $(\hat{\iota}^{EV}_\text{in}, \hat{\iota}^{EV}_\text{out})$,
$\hat\alpha^{EV}$ is calculated by \eqref{eq:alpha:solve} using $\ntail=200$. 


As seen here, for simulated data from a known model, MLE outperforms
other estimation procedures.  The EV procedure tends to have much
larger variance than both MLE and SN with slightly more bias.  This is
not surprising as the performance of the EV estimators is dependent on
the \new{quality of the following approximations:}
\begin{enumerate}
\item The \wtd{number of edges in the} network, $n$,
should be sufficiently large to ensure a close approximation of 
$N_n(i,j)/N(n)$ to the
limit \wtd{joint pmf} $p_{ij}$.
\item  The choice of thresholds must guarantee the quality of the EV
estimates for the indices and the limiting angular distribution.
The thresholding means estimates are based on only a small fraction of the data
and hence have large uncertainty.
\item The parameter $a$ used to
transform the in-  and out-degrees to standard regular variation
is estimated and thus subject to estimation error which propagates
throughout the remaining estimation procedures.
 \end{enumerate}

 \subsubsection{Sensitivity analysis.}\label{subsubsec:sens}

We {explore} {how sensitive {EV estimates are} to choice of $r$}, the threshold for the approximation to the limiting angular density in \eqref{eq:alpha:solve}. Equivalently, we consider varying $n_\text{tail}$, the number of tail observations included in \phy{the} estimation. 

\begin{figure}[t]
	\centering
	\includegraphics[scale=0.6]{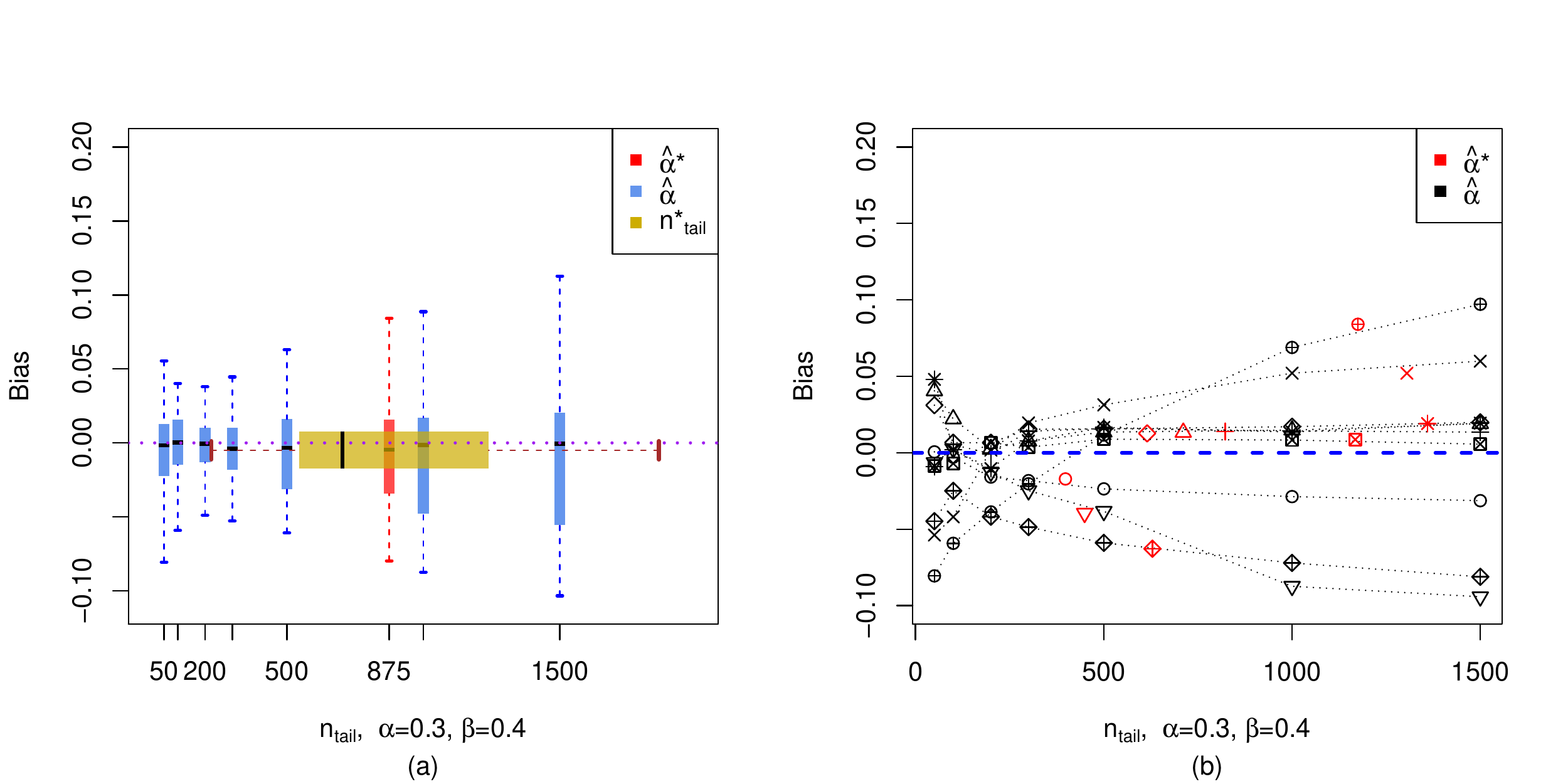}
	\caption{(a) Boxplots of biases of $\hat\alpha$ and $\hat{\alpha}^*$ for different $n_\text{tail}$ and $n_\text{tail}^*$ over 50 replications, where $(\alpha,\beta,\gamma,\din,\dout)=(0.3,0.4,0.3,1,1)$. (b) Linearly interpolated trajectories of biases of $\hat\alpha$ and $\hat{\alpha}^*$ from 10 randomly picked realizations.}\label{fig:alpha_est}
\end{figure}

For the sensitivity analysis, 50 linear PA networks with $10^5$ edges and parameter set
$$(\alpha,\beta,\gamma,\din,\dout)=(0.3,0.4,0.3,1,1),$$
or equivalently,
$$(\alpha,\beta,\gamma,\ain,\aout)=(0.3,0.4,0.3,2.29,2.29)$$
are generated.  We use
$n_\text{tail} = 50,100,200,300,500,1000, 1500$ to calculate the
EV estimates for $\alpha$.
The performances of $\hat{\alpha}^{EV}$ across different value\wtd{s} of $n_\text{tail}$ are demonstrated by the blue boxplots in Figure~\ref{fig:alpha_est}(a).
We see that the biases of $\hat{\alpha}$ remain small until $n_\text{tail}$
increases to 300, and for larger values of $n_\text{tail}$,
$\hat{\alpha}$ considerably underestimates $\alpha$. 

We note that the angular components $R_n(v)$, $1\le v\le N(n)$ are
also power-lawed. As an attempt to select the optimal value of
$n_\text{tail}$, we apply the minimum distance method to the
$R_n(v)$'s and use the selected threshold, $n^*_\text{tail}$, as the
truncation threshold. The boxplot of $n^*_\text{tail}$ for the 50
simulated networks are represented by the \phy{horizontal} boxplot in
Figure~\ref{fig:alpha_est}(a).  The EV estimator with respect to this
threshold for each simulation, denoted by $\hat{\alpha}^*$, is shown
by the red \wtd{boxplot} and plotted at $n_\text{tail}=875$, the mean of
$n^*_\text{tail}$. Overall, $n^*_\text{tail}$ varies between 300 and
1500 and results in an underestimated  
$\hat{\alpha}^*$. 

In Figure~\ref{fig:alpha_est}(b), we randomly choose 10 realizations (among the 50 replications) and plot the linearly interpolated trajectories of $\hat{\alpha}$, based on different values of $n_\text{tail}$. Black points are the estimation results using fixed thresholds $n_\text{tail} = 50,100,200,300,500,1000,1500$ and red ones are determined by $(\hat{\alpha}^*,n^*_\text{tail})$ using the minimum distance method.
Black and red points denoted by the same symbol belong to the same realization.
Comparison among estimation results for
different values of $n_\text{tail}$ reveals that choosing a fixed threshold $n_\text{tail}\le 300$ outperforms selecting a $n_\text{tail}^*$ using the minimum distance method, as it produces estimates with smaller biases and variances. 

\subsection{Data {corrupted} by {random edge addition/deletion}.}\label{sec:perturb}
PA models are designed to describe human
interaction in social networks
but what if data collected from a network is corrupted or usual
behavior is changed? Corruption could be 
due to collection error and atypical behavior could result from
users hiding their network presence or trolls acting as provocateurs.
In such circumstances,
the task is to unmask data corruption or atypical behavior and 
recover {the parameters associated with the
original preferential attachment rules. 

In the following, we {consider network data that are generated from the linear PA model but \wtd{corrupted} by random addition or deletion of edges}. 
{For such corrupted data}, {we attempt to recover the original model and compare the performances of MLE, SN, and EV methods.}

\subsubsection{Randomly adding edges.}\label{subsec:add}

We consider a network generating algorithm with linear PA rules but also a possibility of adding random edges.  
Let $G(n)=(V(n),E(n))$ denote the graph at time $n$. We assume
that the edge set $E(n)$ can be decomposed into two disjoint subsets:
$E(n) = E^{PA}(n) \bigcup E^{RA}(n)$, where $E^{PA}(n)$ is the set of edges
result{ing} from \wtd{PA rules}, and $E^{RA}(n)$ is the set
of those result{ing} from random attachment{s}. {This can be viewed as} an interpolation of
the \wtd{PA} network and the Erd\"os-R\'enyi random
graph.

{More specifically, consider the following network growth}. Given $G(n-1)$, $G(n)$ is formed by creating a new edge where:
\begin{enumerate}
\item[(1)]
	With probability $p_a$, two nodes are chosen randomly
        {(allowing repetition)} from $V(n-1)$ and an edge is created connecting them. The possibility of a self loop is allowed.
\item[(2)]
	With probability $1-p_a$, a new edge is created according to the preferential attachment scheme $(\alpha,\beta,\gamma,\din,\dout)$ on $G^{PA}(n-1):=(V(n-1),E^{PA}(n-1))$.
\end{enumerate}

{The question of interest is,} {if we {are unaware of the perturbation effect and} pretend the data from this model {are} coming from the linear PA model, can}
 we recover the \wtd{PA} parameters?
{To investigate,} we generate networks of $n=10^5$ edges with parameter values
 $$(\alpha,\beta,\gamma,\din,\dout)=(0.3,0.4,0.3,1,1), \quad p_a\in\{0.025,0.05,0.075,0.1,0.125,0.15\}.$$
For each network, the original \wtd{PA} model {is
  fitted} using the {MLE, SN and EV} methods, respectively. The
{angular} MLE  in \eqref{eq:alpha:solve}
in the {extreme value} estimation {is performed} based
on $\ntail=500$ tail observations. In order to compare these estimators, we 
repeat the experiment 200 times for each value of $p_a$ and obtain 200
sets of estimated parameters for each method.  
{Figure~\ref{fig:add_params} {summarizes the estimated values for $(\din, \dout, \alpha, \gamma, \ain, \aout)$ for different values of $p_a$. The mean estimates are marked by crosses and the $2.5\%$ and $97.5\%$ empirical quantiles are marked by the bars.}
The true value of parameters are {shown} as {the} horizontal lines. }

{While all parameters deviate} from the true value as $p_a$ increases
{and} the network becomes more ``noisy", {the EV estimates for
  $(\din,\dout)$ exhibit  smaller bias than the MLE and \wtd{SN}
  methods (Figure \ref{fig:add_params} (a) 
and (b)).} All three methods give underestimated probabilities
$(\alpha, \gamma)$ (Figure \ref{fig:add_params} (c) 
and (d)). This is because the perturbation step (1) creates more edges between existing nodes and consequently inflates the estimated value of $\beta$.

{Also note that the mean {EV} estimates of $(\ain,  
\aout)$ stay close to the theoretical
values for all choices of $p_a$; see Figure \ref{fig:add_params} (e)
and (f). {The MLE and SN estimates of $(\ain,\aout)$, which are computed from the corresponding estimates for $(\alpha,\beta,\gamma,\deltain,\deltaout)$, show strong bias as $p_a$ increases.} 
{In this case}, the \wtd{EV} method is {robust for estimating
  the \wtd{PA} parameters} {and recovering} the tail
indices from the original model.

\begin{figure}[t]
	\centering
	\includegraphics[scale=0.6]{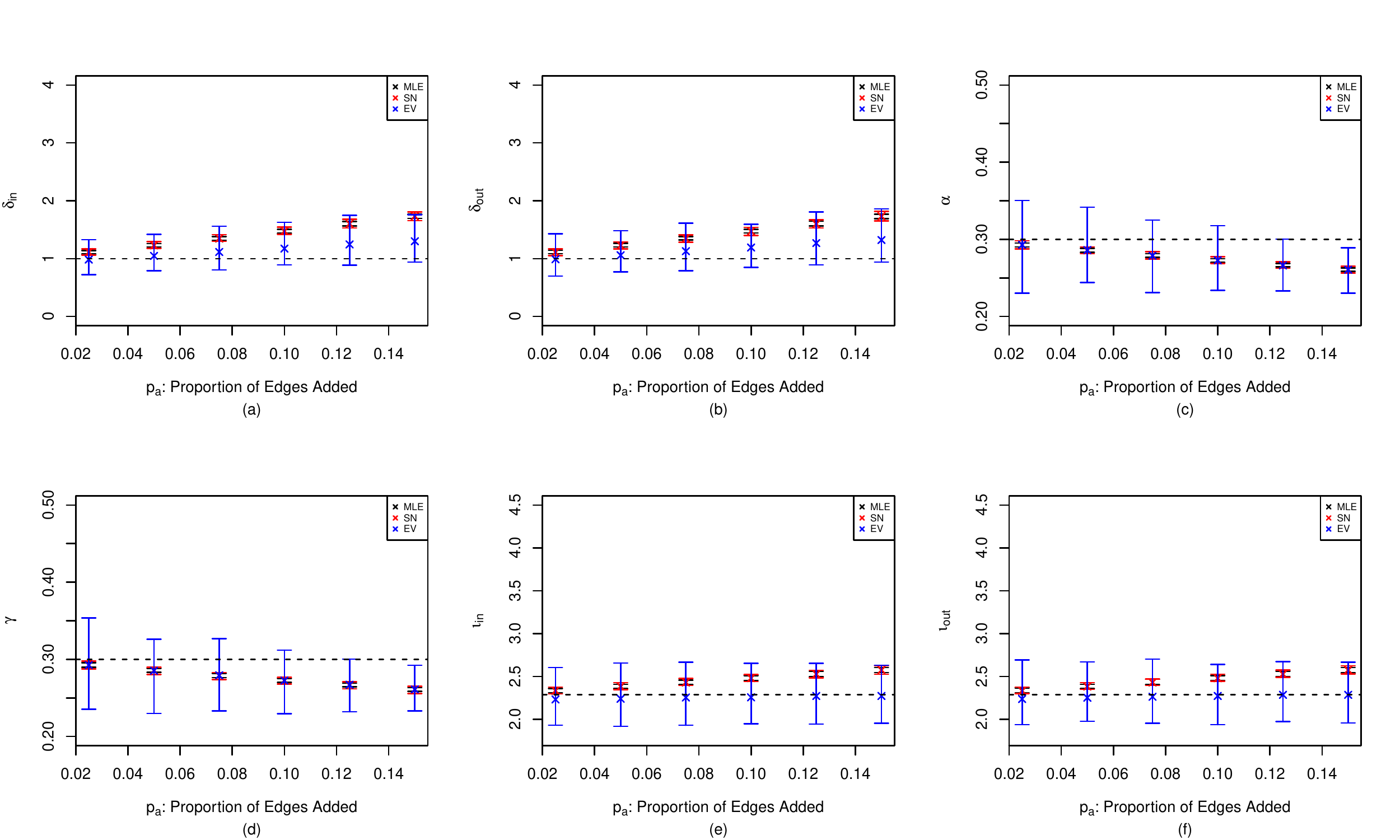}
	\caption{Mean estimates and $2.5\%$  and $97.5\%$  empirical quantiles of (a) $\din$; (b) $\dout$; (c) $\alpha$; (d) $\gamma$; (e) $\ain$; (f) $\aout$, using
	MLE (black), SN (red) and EV (blue) methods over 200 replications, 
	 where $(\alpha,\beta,\gamma,\din,\dout)=(0.3,0.4,0.3,1,1)$ and $p_a= 0.025,0.05,0.075,0.1,0.125,0.15$.
	 For the EV method, 500 tail observations were used {to obtain $\hat{\alpha}^{EV}$}.  }\label{fig:add_params}
\end{figure}

\subsubsection{Randomly deleting edges.}\label{subsec:delete}
{We now consider the scenario where a network is generated from the
  linear PA model, but a random proportion $p_d$ of edges are deleted
  at the final time.  We do this by generating $G(n)$ and then
  deleting $[np_d]$ edges by sampling without replacement.
For the simulation, we generated networks
  with parameter values 
$$(\alpha,\beta,\gamma,\din,\dout)=(0.3,0.4,0.3,1,1), \quad p_d\in\{0.025,0.05,0.075,0.1,0.125,0.15\}.$$
{Again, for each value of $p_d$, the experiment is repeated 200 times and the resulting parameter plots are shown in} Figure~\ref{fig:delete_params} {using the same format as for Figure~\ref{fig:add_params}}. For the \wtd{EV} method, 100 tail observations were used {to {compute} an $\hat{\alpha}^{EV}$}.

{Surprisingly, for all six parameters considered}, MLE estimates stay almost unchanged for different values of $p_d$ while SN and EV estimates underestimate $(\din,\dout)$ and overestimate $(\alpha, \gamma)$, with increasing magnitudes of biases as $p_d$ increases. For tail estimates, the minimum distance method still gives reasonable results (though with larger variances), whereas the \wtd{SN} method keeps underestimating $\ain$ and $\aout$.

The performance of MLE in this case is surprisingly competitive. This
is intriguing and in ongoing work, we will think about why this is the case.

\begin{figure}[t]
	\centering
	\includegraphics[scale=0.6]{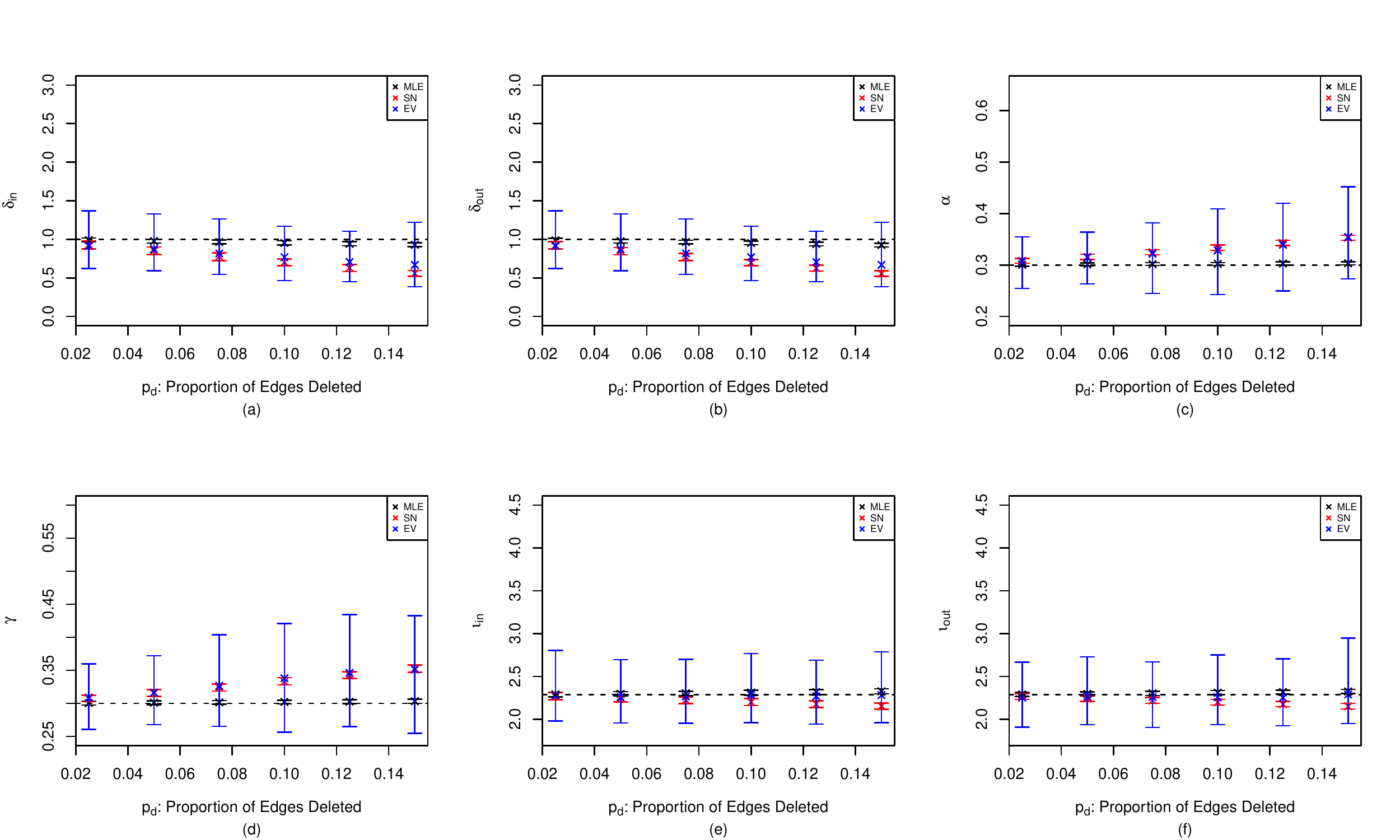}
	\caption{Mean estimates and $2.5\%$  and $97.5\%$  empirical quantiles  of (a) $\din$; (b) $\dout$; (c) $\alpha$; (d) $\gamma$; (e) $\ain$; (f) $\aout$, using
	MLE (black), SN (red) and EV (blue) methods over 50 replications, 
	 where $(\alpha,\beta,\gamma,\din,\dout)=(0.3,0.4,0.3,1,1)$ and $p_d= 0.025,0.05,0.075,0.1,0.125,0.15$.
	 For the EV method, 100 tail observations were used {to compute $\hat{\alpha}^{EV}$}.}\label{fig:delete_params}
\end{figure}

\subsection{Superstar model.}\label{subsec:superstar}
In this section, we consider network data generated from the superstar model.  We compare the accuracy of tail index estimates under parametric methods applied to the linear PA model with extreme value estimates applied directly to data.

{Networks are simulated from the superstar model with the following parameter values:}
$$(\alpha, \beta, \din, \dout, n)= (0.3, 0.4, 0.3, 1, 1, 10^6), \quad p \in \{0.1, 0.15, 0.2, 0.25, 0.3\}.$$
The MLE estimates of the tail indices based on {\eqref{c1c2}}, $(\hat{\iota}^{MLE}_\text{in},\hat{\iota}^{MLE}_\text{out})$, are compared to the EV estimates calculated directly from the node degree data, $(\hat{\iota}^{EV}_\text{in},\hat{\iota}^{EV}_\text{out})$.
According to Theorem~\ref{thm:superstar}, {the} theoretical marginal tail indices for
$I_n(v)$ and $O_n(v)$, \wtd{$1\le v\le N(n)$}, based on a superstar \wtd{PA} model are
{given by \eqref{iotain}, \eqref{iotaout}.}
This experiment is repeated 50 times and Table~\ref{robust-clauset} records the mean estimates for $(\ain, \aout)$ over these 50 replications.

\begin{table}[h]
\centering
\begin{tabular}{ccccc}
 \hline
 $p$& $(\ain, \aout)$ & $(\hat{\iota}^{MLE}_\text{in},\hat{\iota}^{MLE}_\text{out})$ & 
 $(\hatain^{EV},\hataout^{EV})$ \\
 \hline
  $0.1$ & (2.43, 2.29) & (2.11, 2.31) & (2.24 2.25)\\
  $0.15$ & (2.51, 2.29) & (2.03, 2.33) & (2.28 2.20) \\
    $0.2$ & (2.61, 2.29) & (1.97, 2.34) & (2.35 2.18)\\
  $0.25$ & (2.71, 2.29) & (1.91, 2.36) & (2.43 2.18) \\
  $0.3$ & (2.84, 2.29) & (1.86, 2.38) & (2.51 2.15)\\
   \hline
\end{tabular}
\caption{Mean estimates for $(\ain, \aout)$ using both MLE and minimum
  distance  methods, with $(\alpha, \beta, \gamma, \din, \dout, n) =
  (0.3, 0.4, 0.3, 1, 1, 10^6)$.}\label{robust-clauset} 
\end{table}

As $p$ increases and the influence of the superstar node
becomes more profound, the MLE method does not give an accurate 
estimate of tail indices, while the {EV} method stays more robust. However, when $p$ becomes too large, the in-degrees 
of non-superstar nodes will be greatly restricted, which increases the finite sample bias in the {EV} estimates.

Note that the theoretical indices $(\ain, \aout)$ in Table~\ref{robust-clauset} are for the in- and out-degrees of the non-superstar nodes.  In the EV methods, the inclusion of the superstar node \phy{can} severely \phy{bias} the estimation of $\ain$. Let $k_n$ be some intermediate sequence such that 
$k_n\to\infty$ and $k_n/n\to 0$ as $n\to\infty$ and use $I_{(1)}\ge\ldots\ge I_{(k_n+1)}$ to denote the upper $k_n+1$ order statistics of 
$\{I_n(v): 0\le v\le N(n)\}$. Then the corresponding Hill estimator is
\begin{align}
\new{1/\hatain^{EV} (k_n)} &:= \frac{1}{k_n}\sum_{i=1}^{k_n} \log \frac{I_{(i)}}{I_{(k_n+1)}}\nonumber\\
&= \frac{1}{k_n}\log I_{(1)} - \frac{1}{k_n}\log I_{(k_n+1)} + \frac{1}{k_n}\sum_{i=2}^{k_n} \log \frac{I_{(i)}}{I_{(k_n+1)}}.
\label{eq:hill_superstar}
\end{align}
From the construction of the superstar model, we know that the
superstar node 
\new{likely has} the largest in-degree, which is
approximately equal to 
$np$ for large $n$. Hence, the first term in \eqref{eq:hill_superstar}
goes to 0, as long as 
\[
k_n/\log n\to \new{\infty},\quad\text{as }n\to\infty,
\]  
and the third term in \eqref{eq:hill_superstar} is the Hill estimator computed from the in-degrees of non-superstar nodes.
In \cite{wang:resnick:2017}, the consistency of the Hill estimator has
been proved for a simple undirected linear PA model, but consistency for
 $\hatain^{EV}(k_n)$ is not proven for either  of the two models we consider here. 
However, with the belief on the consistency of $\hatain^{EV}(k_n)$, \eqref{eq:hill_superstar} suggests that choosing a larger $k_n$ will 
reduce the bias when estimating $\ain$ in the superstar model.  

To illustrate this point numerically, we choose $k_n = 200, 500, 1000, 1500, 2000$ 
for a superstar network with $10^6$ edges and 
probability of attaching to the superstar node $p = 0.1, 0.15, 0.2, 0.25, 0.3$.
For each value of $p$, we again simulate 50 independent replications of the superstar PA model with parameters
$(\alpha, \beta, \gamma, \din, \dout, n) = (0.3, 0.4, 0.3, 1, 1, 10^6)$. Then for each replication generated, 
Hill estimates of the in- and out-degree tail indices are calculated under different choices of $k_n$. The mean values of the 
50 pairs of estimates are recorded in Table~\ref{Table:vary_kn}, where the first entry is the in-degree tail estimate and the second is for out-degree.

\begin{table}[h]
\centering
\begin{tabular}{l|ccccc}
\hline
& \multicolumn{5}{c}{Number of Upper Order Statistics $k_n$} \\
 & 200 & 500 & 1000 & 1500 & 2000\\ 
\hline 
$p = 0.1$ & (2.16, 2.22) & (2.26, 2.19) & (2.27, 2.16) & (2.28,  2.14) & (2.27, 2.15) \\ 
$p = 0.15$ & (2.25, 2.18) & (2.32, 2.17) & (2.29, 2.14) & (2.31,2.15) & (2.28, 2.14)\\ 
$p = 0.2$ & (2.32, 2.17) & (2.39, 2.16) & (2.37, 2.15) & (2.39, 2.11) & (2.33, 2.13) \\ 
$p = 0.25$ & (2.36, 2.18) & (2.47, 2.16) & (2.43, 2.12) & (2.49, 2.11) & (2.52, 2.12)\\ 
$p = 0.3$ & (2.41, 2.17) & (2.58, 2.13) & (2.56, 2.11) & (2.47, 2.11) & (2.51, 2.12)\\ 
\hline
\end{tabular} 
\caption{Mean values of \new{EV} 
estimates of tail indices $(\ain, \aout)$ over 50 replications,
 with $(\alpha, \beta, \gamma, \din, \dout, n) = (0.3, 0.4, 0.3, 1, 1,
 10^6)$.
\new{The true values are given in Table \ref{robust-clauset}.}} 
\label{Table:vary_kn}
\end{table}
From the in-degree estimates in Table~\ref{Table:vary_kn}, we observe that for most values of $p$ increasing $k_n$ to 500
improves the estimation results, but further increase in $k_n$ has
adverse effects. One reason is that \new{large $k_n$} means
smaller in-degrees are taken into  
the calculation of the Hill estimator; these smaller in-degrees
 might not be large enough to be considered as following the power law
 in \eqref{power-in}. 
This also explains the increasing biases for the out-degree estimates, where the superstar node does not have any impact.
Comparing the results in Table~\ref{Table:vary_kn} to those EV estimates in Table~\ref{robust-clauset}, we see
that the minimum distance method seeks a good balance between eliminating the effect of the superstar nodes and choosing a reasonably 
large threshold.

\begin{figure}
\includegraphics[scale = 0.65]{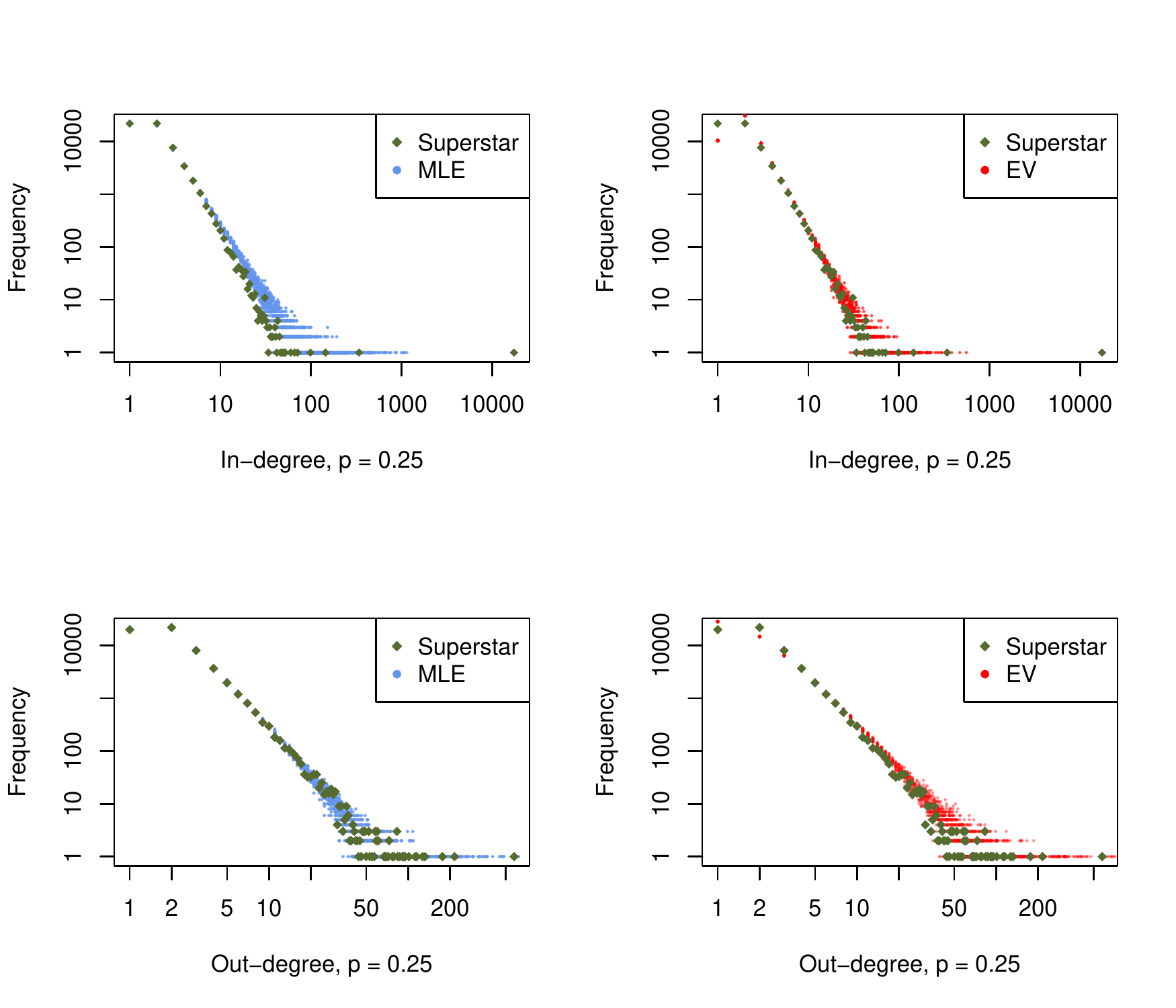}
\caption{Empirical in- and out-degree distributions, with $(\alpha, \beta, \gamma, \din, \dout,n, p) = (0.3,\, 0.4,\, 0.3,\, 1,\, 1,\, 10^5,\, 0.25)$.}\label{fig:deg_compare}
\end{figure}

The next question is how the model misspecification affects the empirical
  distributions of in- and out-degrees.
To {evaluate} this, we generated a superstar \wtd{PA} model with parameters
$$
(\alpha, \beta, \gamma, \din, \dout,n, p) = (0.3,\, 0.4,\, 0.3,\, 1,\, 1,\, 10^5,\, 0.25). 
$$
We {estimated parameters by}
 both MLE {and {EV} methods} {from} simulated
 superstar {data}, pretending that the data was generated from an ordinary
\wtd{PA} rule. For the
\wtd{EV} approach, 200 tail observations \new{were}
 used {while computing $\hat{\alpha}^{EV}$}.
Denote the MLE and \wtd{EV} estimates by 
\begin{align*}
\widehat{\boldsymbol\theta}_n^{MLE} &:= (\hat{\alpha}^{MLE}, \hat{\beta}^{MLE}, \hat{\gamma}^{MLE}, \hat{\delta}^{MLE}_\text{in}, \hat{\delta}^{MLE}_\text{out}), \\
\widehat{\boldsymbol\theta}_n^{EV} &:= (\hat{\alpha}^{EV},
                                      \hat{\beta}^{EV},
                                      \hat{\gamma}^{EV},
                                      \hat{\delta}^{EV}_\text{in},
                                      \hat{\delta}^{EV}_\text{out}). 
\end{align*}
We then simulated $20$ independent replications of a linear
PA model with parameters
$\widehat{\boldsymbol\theta}_n^{MLE}$ and $20$ with parameters 
$\widehat{\boldsymbol\theta}_n^{EV}.$   For each set of replicates we
computed the empirical frequency distributions.
Comparisons of degree distributions are provided in
Figure~\ref{fig:deg_compare}.

In all 4 panels, the green dots represent the empirical degree frequencies for
the simulated superstar data, top for in-degree and bottom
for out-degree. Blue in the two left panels
represents overlaid frequency distributions for the 20 simulated data
sets from the \wtd{linear PA} replicates using 
$\widehat{\boldsymbol\theta}_n^{MLE}$. Red in the right two panels
does the same thing for 20 replicates of the \wtd{linear PA}
model using parameter $\widehat{\boldsymbol\theta}_n^{EV}$.

The \wtd{EV} method {seems} to give better fit for
in-degrees. Based on out-degrees,  it is difficult to visually discern
an advantage for either approach.  {While not obvious in the plots, we again expect the estimated degrees from the EV method to have higher variance than those from MLE, as much less data were used for the model fitting.}

\section{Conclusion}\label{sec:discussion}

{In this paper, we propose{d} a semi-parametric {extreme value} (EV)
  estimation method for network models.  We compared the {performance}
  of this method to the two parametric approaches (MLE and snapshot
  methods) given in \cite{wan:wang:davis:resnick:2017} under three
  scenarios: (1) data generated from a linear preferential attachment
  (linear PA) model;  
(2) data generated from a linear PA model with corruption; (3) data generated from a superstar linear PA model.}

To summarize our findings and experience, EV
estimation methods play important roles while applied to social
network data.  The method provides a robust procedure for
estimating \new{parameters} of the network related to heavy-tailedness of
the marginal and joint distributions of the in- and out-degrees.
Also EV methods  play a confirmatory role to other
estimation procedures that are likelihood based, such as MLE or the
snapshot (SN) method, which require that the model is correctly
specified.  If, for example, MLE or SN produces estimates of tail
indices different \wtd{from} those \wtd{given} by the EV procedure, then this might
suggest a lack of fit of the underlying model.  

In practice, data are
not as {\it clean} as those produced in simulations and one
expects deviations from a base model such as the linear PA.
As seen in this paper, these deviations can lead to sharply biased MLE
and SN estimates especially when compared to EV estimates.  As in
classical \wtd{EV} estimation in the iid setting, the
choice of threshold upon which to base the estimation remains a thorny
issue in the network context.  The minimum distance method based on
\cite{clauset:shalizi:newman:2009} for estimating 
\new{marginal} tail indices works
well for the examples considered here, but \wtd{worse} for multivariate
data where it is employed to set thresholds based on radius vectors.

\appendix

\section{Parameter Estimation for linear PA Model}\label{subsec:param_est} 
Parameter estimation {for the linear PA model was studied} in \cite{wan:wang:davis:resnick:2017}. 
{If} the complete history of {the} network evolution {is} available ({i.e.}, {timestamps of edge creation are known), then} MLE estimates {exist and are computable}.
{On the other hand, if only a snapshot of the network is given at a single point in time (i.e., timestamp information for the creation of the edges is unavailable)}, {an approximate MLE was proposed}.  {This procedure combined elements of method of moments with an approximation to the likelihood.} {In the following} we {provide} a brief summary of these two estimation methods.  {Asymptotic properties of these estimators can be found in \cite{wan:wang:davis:resnick:2017}.} 
\subsection{MLE}\label{subsub:mle}
{Given the full evolution of the network $G(n)$, assuming the graph began with $n_0$ initial edges, the MLE estimator of $\boldsymbol\theta = (\alpha,\beta,\gamma,\din,\dout)$,
$$\widehat{\boldsymbol{\theta}}_n^{MLE}:=(\hat{\alpha}^{MLE}, \hat{\beta}^{MLE}, \hat{\gamma}^{MLE}, \hatdin^{MLE},\hatdout^{MLE}),$$
{is obtained} by setting
\begin{align*}
\hat{\alpha}^{MLE} &= \frac{1}{n-n_0} \sum_{t=n_0+1}^n \ind_{\{J_t=1\}},\\
\hat{\beta}^{MLE} &= \frac{1}{n-n_0} \sum_{t=n_0+1}^n \ind_{\{J_t=2\}},\\
\hat{\gamma}^{MLE} &= 1- \hat{\alpha}^{MLE} - \hat{\beta}^{MLE},\\
\intertext{and solving for $(\hatdin^{MLE}, \hatdout^{MLE})$ from}
\sum_{i=0}^\infty \frac{\Nin_{>i}(n) - \Nin_{>i}(n_0)}{i+\hatdin^{MLE}}
&= \frac{n-n_0}{\hatdin^{MLE}}\hat{\gamma}^{MLE}
+\sum_{t=n_0+1}^n \frac{N(t-1)}{t-1+\hatdin^{MLE} N(t-1)}\ind_{\{J_t\in\lbrace 1, 2\rbrace\}},\\
\sum_{j=0}^\infty \frac{\Nout_{>j}(n) - \Nout_{>j}(n_0)}{j+\hatdout^{MLE}}
&= \frac{n-n_0}{\hatdout^{MLE}}\hat{\alpha}^{MLE}
+\sum_{t=n_0+1}^n \frac{N(t-1)}{t-1+\hatdout^{MLE} N(t-1)}\ind_{\{J_t\in\lbrace 2, 3\rbrace\}},
\end{align*}
where 
\[
\Nin_{>i}(n) := \sum_{i'>i} \Nin_{i'}(n),\qquad \Nout_{>j}(n) := \sum_{j'>j} \Nout_{j'}(n).
\]
By \cite[Theorem~3.3]{wan:wang:davis:resnick:2017}, $\widehat{\boldsymbol{\theta}}_n^{MLE}$ {is} strongly consistent, asymptotically normal and efficient.
\subsection{Snapshot.}
{The} estimation method for $\boldsymbol \theta$ from the snapshot $G(n)$ {is summarized in} the following 7-step procedure:
\begin{enumerate}
\item[1.] {Estimate $\beta$} by $\hat{\beta}^{SN}=1-N(n)/n$.
\item[2.] Obtain $\hatdin^0$ by solving 
$$
\sum_{i=1}^\infty \frac{\Nin_{>i}(n)}{n}\frac{i}{i+\hatdin^0}(1+\hatdin^0(1-\hat{\beta}^{SN})) =\frac{\frac{\Nin_0(n)}{n} + \hat{\beta}^{SN} }{1-\frac{\Nin_0(n)}{n} \frac{\hatdin^0}{1+(1-\hat{\beta}^{SN})\hatdin^0}},
$$
where $\Nin_0(n)$ denotes the number of nodes with in-degree 0 in $G(n)$.
\item[3.] Estimate $\alpha$ by
$$
\hat\alpha^0 =  \frac{\frac{\Nin_0(n)}{n} +\hat{\beta}^{SN}}{1-\frac{\Nin_0(n)}{n} \frac{\hatdin^0}{1+(1-\hat{\beta}^{SN})\hatdin^0}} - \hat{\beta}^{SN}.
$$
\item[4.] Obtain $\hatdout^0$ by solving
$$
\sum_{j=1}^\infty \frac{\Nout_{>j}(n)}{n}\frac{j}{j+\hatdout^0}(1+\hatdout^0(1-\hat{\beta}^{SN})) = \frac{\frac{\Nout_0(n)}{n} + \hat{\beta}^{SN} }{1-\frac{\Nout_0(n)}{n} \frac{\hatdout^0}{1+(1-\hat{\beta}^{SN})\hatdout^0}},
$$
where $\Nout_0(n)$ denotes the number of nodes with out-degree 0 in $G(n)$.
\item[5.] Estimate $\gamma$ by
$$
\hat\gamma^0 =  \frac{\frac{\Nout_0(n)}{n} + \hat{\beta}^{SN}}{1-\frac{\Nout_0(n)}{n} \frac{\hatdout^0}{1+(1-\hat{\beta}^{SN})\hatdout^0}} - \hat{\beta}^{SN}.
$$
\item[6.]  Re-normalize the probabilities
$$ \label{eq:renormalize}
	(\hat\alpha^{SN},\hat{\beta}^{SN},\hat\gamma^{SN}) \leftarrow \left(\frac{\hat\alpha^0(1-\hat{\beta}^{SN})}{\hat\alpha^0+\hat\gamma^0},\hat{\beta}^{SN},\frac{\hat\gamma^0(1-\hat{\beta}^{SN})}{\hat\alpha^0+\hat\gamma^0}\right).
$$
\item[7.] Solve for $\hatdin^{SN}$ from
$$
	\sum_{i=0}^\infty \frac{\Nin_{>i}(n)/n}{i+\hatdin^{SN}}-\frac{1-\hat\alpha^{SN}-\hat{\beta}^{SN}}{\hatdin^{SN}}
-\frac{(\hat\alpha^{SN}+\hat{\beta}^{SN})(1-\hat{\beta}^{SN})}{1+(1-\hat{\beta}^{SN})\hatdin^{SN}} = 0.
$$
Similarly, solve for $\hatdout^{SN}$ from
$$
	\sum_{j=0}^\infty \frac{\Nout_{>j}(n)/n}{j+\hatdout^{SN}}-\frac{1-\hat\gamma^{SN}-\hat{\beta}^{SN}}{\hatdout^{SN}}
-\frac{(\hat\gamma^{SN}+\hat{\beta}^{SN})(1-\hat{\beta}^{SN})}{1+(1-\hat{\beta}^{SN})\hatdout^{SN}} = 0.
$$
\end{enumerate}
\smallskip
{Note that Step 6 ensures that
$$ \label{eq:probeq}
	\hat\alpha^{SN}+\hat{\beta}^{SN}+\hat\gamma^{SN}=1.
$$
It is shown in \cite[Theorem~4.1]{wan:wang:davis:resnick:2017} that  $\widehat{\boldsymbol \theta}^{SN}_n := (\hat\alpha^{SN},\hat\beta^{SN},\hat\gamma^{SN}, \hatdin^{SN}, \hatdout^{SN})\convas \boldsymbol \theta$.}  Its asymptotic normality and efficiency are 
analyzed through simulation studies in {the same paper}.


\section{Proof of Theorem~\ref{thm:superstar}}\label{subsec:proof:superstar}
\begin{proof}
We first prove the out-degree part of Theorem~\ref{thm:superstar}.
Note that 
\begin{align}\label{eq:out_deg}
\EE\left(\Nout_j(n+1)| G(n)\right)& =\Nout_j(n)+\gamma\boldsymbol{1}_{\{j=0\}}+\alpha\boldsymbol{1}_{\{j=1\}}\nonumber\\
 +& (\beta+\gamma)
\left(\Nout_{j-1}(n)\frac{j-1+\dout}{n+\dout|V^0(n)|} - \Nout_{j}(n)\frac{j+\dout}{n+\dout|V^0(n)|}\right).
\end{align}
Meanwhile, by the definition of $V^0(n)$, we have
\beqq\label{V0_dist}
|V^0(n)|+1 = N(n) \sim \text{Binomial}(n, 1-\beta).
\eeqq

Applying the arguments in the proof of Theorem~3.1 of
\cite{bollobas:borgs:chayes:riordan:2003}, it follows that the
out-degree distribution of a linear superstar model coincides with
that of a standard linear preferential attachment network 
 with parameters $(\alpha,\beta,\gamma,\din,\dout)$.
Moreover,
$$
	\frac{\Nout_j(n) }{n}  \convas \qout_j, \quad j>0,\quad n\to\infty,
$$
where $\{\qout_j\}:=\{\pout_j\}$ is the limiting out-degree distribution of PA$(\alpha,\beta,\gamma,\din,\dout)$. In particular,
$$
	\qout_j \, \sim \Cout'\, j^{-(1+\iota_{\text{out}})}\qquad\mbox{ as }j\to\infty, 
$$
for $\Cout'$ positive and 
$$
	\iota_{\text{out}}^{-1} \,=\, \frac{\beta+\gamma}{1+\dout(\alpha+\gamma)}.
$$

Next we consider the in-degree counts of non-superstar nodes. Observe also from the construction of the superstar model that 
\beqq\label{E0_dist}
|E^0(n)| \sim \text{Binomial}(n, 1-(\alpha+\beta)p).
\eeqq
Applying the Chernoff bound to both \eqref{V0_dist} and \eqref{E0_dist} gives
\begin{align*}
\left|V^0(n)\right| &= (1-\beta)n + O(n^{1/2}\log n),\\
\left|E^0(n)\right| &= (1-(\alpha+\beta)p)n + O(n^{1/2}\log n).
\end{align*}
Taking expectation on both sides of \eqref{eq:out_deg} then gives
\begin{align}\label{eq:bound}
\EE &\left((\alpha+\beta)(1-p)\Nin_i(n)\frac{i+\din}{|E^0(n)|+\din |V^0(n)|}\right)\nonumber\\
&= (\alpha+\beta)(1-p) \frac{i+\din}{n(1-(\alpha+\beta)p)+\din \cdot n(1-\beta)} \EE(\Nin_i(n)) + O(n^{-1/2}\log n).
\end{align}
By the rule of the superstar model, given $G(n)$, $\Nin_i(n)$ will increase by 1 if either scenario (1b) or (2b) happens and 
a node with $\Din^{(n)}(v) = i-1$ is chosen as the ending point of the edge. Also, it will decrease by 1 if either scenario (1b) or (2b) happens, but 
a node with $\Din^{(n)}(v) = i$ is chosen as the ending point of the edge. Moreover, with probability $\alpha$ a new node 
with in-degree 0 will be added to the graph, and with probability $\gamma$ a new node with in-degree 1 is created in the next step. 
Hence, $\{\Nin_i(n)\}_{n\ge 1}$ satisfies the following:
\begin{align*}\label{eq:recursion_in}
\EE\left(\Nin_i(n+1)|G(n)\right)
 = & \, \Nin_i(n) + \alpha\boldsymbol{1}_{\{i=0\}}+\gamma\boldsymbol{1}_{\{i=1\}}\nonumber\\
&+ (\alpha+\beta)(1-p)\Nin_{i-1}(n)\frac{i-1+\din}{|E^0(n)|+\din |V^0(n)|}\nonumber\\
&- (\alpha+\beta)(1-p)\Nin_i(n)\frac{i+\din}{|E^0(n)|+\din |V^0(n)|}. 
\end{align*}

Now let $\qin_{-1} = 0$, and define $\{\qin_i\}_{i\ge 0}$ by 
\beqq\label{eq:recursion_qin}
\qin_i = \frac{(\alpha+\beta)(1-p)}{1-(\alpha+\beta)p+\din(\alpha+\gamma)} \left((i-1+\din)\qin_{i-1} - (i+\din)\qin_i\right)
+ \alpha\boldsymbol{1}_{\{i=0\}}+\gamma\boldsymbol{1}_{\{i=1\}}.
\eeqq
According to the approximation in \eqref{eq:bound}, we use the same proof technique as in \cite[Theorem~3.1]{bollobas:borgs:chayes:riordan:2003}
to obtain
\[
\frac{\Nin_i(n)}{n}\convas \qin_i,\qquad\text{as}\quad n\to\infty.
\]
Also, solving the recursion in \eqref{eq:recursion_qin} yields
\begin{align}
\qin_0 &= \frac{\alpha}{1+\iota_{\text{in}}^{-1} \din},\label{eq:qin0}\\
\qin_1 &= (1+\din+\iota_{\text{in}})^{-1}\left(\frac{\alpha\din}{1+\iota_{\text{in}}^{-1}\din}+\frac{\gamma}{\iota_{\text{in}}^{-1}}\right),\nonumber\\
\qin_i &= \frac{\Gamma(i+\din)}{\Gamma(i+\din+\iota_{\text{in}}+1)} \frac{\Gamma(2+\din+\iota_{\text{in}})}{\Gamma(1+\din)} \qin_1,
\quad i\ge2,\label{eq:qini}
\end{align}
where 
\[
\iota_{\text{in}}^{-1} := \frac{(\alpha+\beta)(1-p)}{1-(\alpha+\beta)p+\din(\alpha+\gamma)}.
\]
Therefore, applying Stirling's approximation to \eqref{eq:qin0}--\eqref{eq:qini} gives 
$$ \label{eq:power_in}
\qin_i \sim C'_\text{in}\, i^{-(1+\iota_{\text{in}})}, \qquad \text{as }n\to\infty,
$$
for some positive constant $C'_\text{in}$. This completes the proof.
\end{proof}

\bibliography{./bibfile}

\end{document}